\documentclass[a4paper,fleqn,usenatbib,letters]{mnras}

\usepackage{newtxtext,newtxmath}

\usepackage[T1]{fontenc}

\usepackage{graphicx}	
\usepackage{amsmath}	
\usepackage{longtable}
\usepackage{siunitx}
\DeclareSIUnit\year{yr}
\DeclareSIUnit\mag{mag}
\usepackage{tablefootnote}
\usepackage{xspace}
\usepackage{booktabs}
\usepackage{subcaption}
\usepackage[english]{babel}

\numberwithin{equation}{section}
\graphicspath{{./}{figures/}}
\defcitealias{Rose2019}{R19}
\defcitealias{Rose2020}{R20}
\defcitealias{Lee2020}{L20}
\defcitealias{Kang2020}{K20}

\newcommand{\RoseMC}{\citetalias{Rose2019}\xspace}

\newcommand{\Lee}{\citetalias{Lee2020}\xspace}

\newcommand{\E}[1]{\mathbb{E}\left[#1\right]}

\newcommand{\var}[1]{\text{Var}\left[#1\right]}
\newcommand{\slope}{\beta}
\newcommand{\intercept}{\alpha}
\renewcommand{\vec}{\boldsymbol}

\addto\extrasenglish{%
}

\title[SN~Ia luminosity evolution]{Improving Bayesian posterior correlation analysis on Type Ia supernova luminosity evolution}

\author[K. D. Zhang et al.]{
Keto D. Zhang,$^{1}$\thanks{E-mail: keto.zhang@gmail.com}
Yukei S. Murakami,$^{1,2,3}$ 
Benjamin E. Stahl,$^{1,2,4}$
Kishore C. Patra,$^{1,5}$
\newauthor{
and Alexei V. Filippenko$^{1,6,7}$
}
\\
$^{1}$Department of Astronomy, University of California, Berkeley, CA 94720-3411, USA\\
$^{2}$Department of Physics, University of California, Berkeley, CA 94720-7300, USA\\
$^{3}$Google Lick Predoctoral Fellow\\
$^{4}$Marc J. Staley Graduate Fellow\\
$^{5}$Nagaraj-Noll Graduate Fellow \\
$^{6}$Miller Institute for Basic Research in Science, University of California, Berkeley, CA 94720, USA\\
$^{7}$Miller Senior Fellow
}

\date{Accepted XXX. Received YYY; in original form ZZZ}

\pubyear{2020}

\begin{document}

\label{firstpage}
\pagerange{\pageref{firstpage}--\pageref{lastpage}}
\maketitle

\begin{abstract}
    Much of the cosmological utility thus far extracted from Type Ia supernovae (SNe~Ia) relies on the assumption that SN~Ia peak luminosities do not evolve significantly with the age (local or global) of their stellar environments. Two recent studies have provided conflicting results in evaluating the validity of this assumption, with one finding no correlation between Hubble residuals (HR) and stellar environment age, while the other claims a significant correlation. In this {\it Letter} we perform an independent reanalysis that rectifies issues with the statistical methods employed by both of the aforementioned studies. Our analysis follows a principled approach that properly accounts for regression dilution and critically (and unlike both prior studies) utilises the Bayesian-model-produced SN environment age estimates (posterior samples) instead of point estimates. Moreover, the posterior is used as an informative prior in the regression. We find the Pearson correlation between the HR and local (global) age to be in excess of $4\sigma$ ($3\sigma$). Assuming there exists a linear relationship between HR and local (global) age, we find a corresponding slope of $-0.035 \pm 0.007$\,mag\,Gyr$^{-1}$ ($-0.036 \pm 0.007$\,mag\,Gyr$^{-1}$). We encourage further use of our approach to examine HR and host environment correlations, as well as experiments in correcting for luminosity evolution in SN~Ia standardisation.

\end{abstract}




\section{Introduction}
\label{sec:introduction}

The standardisable property \citep[e.g.,][]{Phillips1993} of Type~Ia supernovae (SNe~Ia) played a pivotal role in the discovery of the accelerating expansion of the Universe \citep{Riess1998,Perlmutter1999}. Subsequently, various improvements have been made to reduce biases induced by environmental effects during standardisation. These improvements all serve to reduce the Hubble residuals (HR) --- the difference between observed and best-fit-cosmology predicted distance moduli on the Hubble–Lemaître diagram  \citep{Lampeitl2010,Sullivan2010,Childress2013}, and are typically manifested through correlation statistics between the HR and various SN~Ia environment observables. For example, a correction for host-galaxy stellar mass is now routinely included in cosmological analyses \citep[e.g.,][]{Betoule2014,Scolnic2018}. Another candidate HR correlate is the age of the host's stellar population \citep{Childress2014}, the significance (and even presence) of which has been vigorously debated in recent studies \citep{Rose2019,Kang2020,Rose2020,Lee2020}. 
The presence of such correlation (and a negative trend) indicates that SN Ia luminosities are fainter as their age (and cosmological redshift) increases. With a sufficiently strong trend, the resulting systematic biases may indicate that the apparent excess decrease in distant SN Ia luminosity is not due to the accelerating expansion of the Universe but merely an artefact of the age-luminosity relationship.

In this {\it Letter}, we follow the argument of \citet[][hereafter \Lee]{Lee2020}  against the methodologies of \citet[][hereafter \RoseMC]{Rose2019} when analysing the relation between HR and two measures of the host stellar population age: (1) the local age of the stellar population near each SN, and (2) the global age of the stellar population of the entire host galaxy. We describe the datasets used by both of the aforementioned studies in Section~\ref{sec:data}, suggest a principled approach to using the datasets for inference in Section~\ref{sec:age_posterior_inference}, and reanalyse the correlation (Sec.~\ref{sec:correlation}) and slope (Sec.~\ref{sec:slope}), rectifying issues with the statistical methods found in both studies.



\section{Data}
\label{sec:data}

We collect HR and SN global and local environment age estimates (collectively called ``Age''; separately ``global Age'' and ``local Age'') from Table 1 and the data repository\footnote{\hyperlink{https://doi.org/10.5281/zenodo.3875481}{https://doi.org/10.5281/zenodo.3875481}} of \RoseMC.
The \RoseMC HR dataset is a modification of the set provided by \cite{Campbell2013}, which uses SNe~Ia from the SDSS-II supernova survey \citep{Sako2008}. The modification corrects the HR value for those that are significantly correlated with the SN~Ia stretch parameter. The HR values in the dataset were determined by a Monte Carlo Markov Chain (MCMC) method used by \cite{Campbell2013}. The HR dataset includes only the mean HR and $1\sigma$ standard deviations of the MCMC posterior. Without the MCMC posterior samples, we can only assume that the HR values have Gaussian uncertainties.

\begin{figure}
    \centering
    \includegraphics[width=\columnwidth]{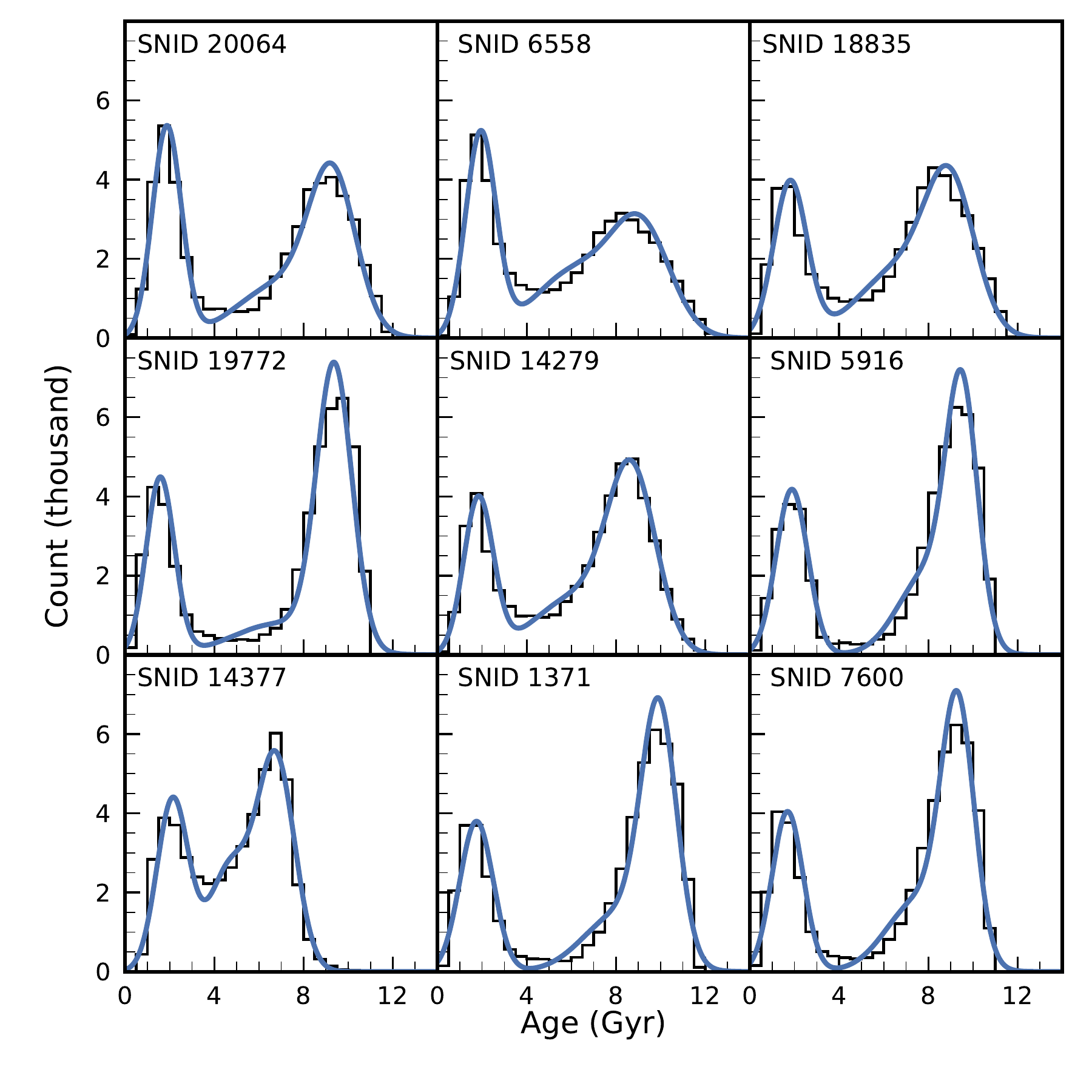}
    \caption{Histograms of the downsampled local Age posterior samples of the first 9 SNe in our selected dataset sorted by highest kurtosis (the set of all 102 Age posterior samples is provided as supplementary data). Strong multimodal features are clearly visible in many of the posterior samples (and as a result, many of these samples are very unlikely to be Gaussian). The blue solid line is the three-component Gaussian mixture model fit to the posterior samples as described in Section~\ref{sec:estimating-the-age-posterior-distribution}.}
    \label{fig:age_posterior_dist}
\end{figure}

Similarly, the \RoseMC Age dataset is derived from an MCMC method. However, the Age dataset includes the entire MCMC posterior sample of size \num{1020000} for each SN. We further discuss this dataset in Section~\ref{sec:age_posterior_inference}. Although \Lee also source their dataset from \RoseMC, they use summary statistical descriptors of the Age dataset (in contrast to the full MCMC posterior samples we use). In particular, it appears that \Lee have retrieved the columns from Table 7 of \RoseMC that correspond to the Age posterior sample's mean, median, standard deviation (SD), $-1$ SD quantile, and $+1$ SD quantile for each SN. We \emph{strongly emphasise} that assembling a dataset in this manner makes assumptions that are incorrect (e.g., modalities in the underlying Age posteriors for each SN, as can be seen in Fig.~\ref{fig:age_posterior_dist}). Consequently, their subsequent estimates of slope and correlation are inherently flawed; we further discuss this in Section~\ref{sec:age_posterior_inference}.

For our analysis, we have removed two SNe that uniquely exist in the HR dataset (SNID 3256) and the Age dataset (SNID 15459). These SNe are also missing in Tables 1 and 7 of \RoseMC, respectively. After removing these two SNe, the resulting dataset comprises a total of 102 SNe.
Owing to the computationally prohibitive size of the Age dataset (\num{1020000} samples $\times$ 102 SNe), we downsample the Age dataset  to \num{50000} for each SN by uniformly sampling \num{50000} rows without replacement for each of the 102 retrieved SNe. We use the Kolmogorov-Smirnoff (KS) test to determine how different the downsample is from the original sample, and we resample until all 102 downsamples have a KS $p$-value greater than 5\% --- there are no significant differences between the downsample and original sample above the $2\sigma$ level.



\section{Age Posterior Inference}
\label{sec:age_posterior_inference}


The Age posterior can be a useful informant in forecasting future observations of the data it initially describes (e.g., the local Age estimates posterior may inform the possible light-curve parameters a particular SN in that environment can have as these two are significantly correlated in a recent study by \cite{Rigault2020}). More applicable to our problem of making statistical inference between the relationship of HR and Age, we may use the Age posterior as a prior to a Bayesian model that estimates the correlation and slope (see Sec.~\ref{sec:correlation} and Sec.~\ref{sec:slope}, respectively). This technique is very similar to hierarchical Bayesian models where priors are conditioned on other priors (which is also true in our case). When we use the Age posterior distribution as a prior distribution in our model, it is conditioned on other priors defined in Equation 9 of \RoseMC. Unlike hierarchical Bayesian models, however, we do not fit for every prior at once, but instead opt for a stepwise process so that we can take advantage of Bayesian simulation results that others have already computed and published.

\subsection{Estimating the Age Posterior Distribution}
\label{sec:estimating-the-age-posterior-distribution}
Although we describe above the predictive power yielded by the posterior distribution, we actually \emph{do not} have the Age posterior distribution. Instead, we have a posterior sample generated by the MCMC sampler under the assumption of the posterior distribution. Fortunately, with a fairly large posterior sample, we can fit for the posterior distribution using a parametric probability density function (PDF).

Examining the various distributions of each SN in \autoref{fig:age_posterior_dist}, we infer that a multimodal distribution better fits the data. Thus, we choose the Gaussian Mixture Model (GMM) as the posterior distribution that fits the Age posterior sample ($A_1, A_2, \ldots, A_N$). We fit the GMM to our Age posterior samples maximising the likelihood probability,

\begin{equation}
    \max_{\vec{\mu}, \vec{\sigma}}{\prod_{i=1}^N\sum_{j=1}^{k} w_j \cdot \text{Gaussian}(A_{i};\, \mu_j, \sigma_j)} \,. \label{eq:max-likelihood}
\end{equation}
\noindent
We set $k=3$ (i.e., three Gaussian components) after observing that all the posterior samples have no more than three significant modes.

\section{Correlation}
\label{sec:correlation}






As expressed above, the estimation of the correlation between HR and Age is independent and separate from the estimation of the slope. Correlation estimation models cannot initially assume that a linear relationship exists between HR and Age while slope estimation models (i.e., linear models) do. Here, we use the Pearson correlation coefficient, $r$, to gauge how strongly two variables are linearly related. The correlation coefficient with uncertainties in both variables (denoted as $\sigma_x$ and $\sigma_y$) can be biased by a relation that is inversely proportional to these uncertainties,

\begin{equation}
    r_{x,y} = \frac{\text{Cov}[x^*, y^*] + \sigma_{x, y}^2}{\sqrt{\left(\var{x^*} + \sigma_x^2 \right) \left(\var{y^*} + \sigma_y^2\right)}} \,, \label{eq:correlation-biased}
\end{equation}

\noindent where the asterisks denote true variables (e.g., $x = x^* + \text{error}$) and $\sigma_{x, y}^2$ is the covariance between the uncertainties (often set to zero by invoking the classical assumption of independent measurement error; we invoke the same assumption hereafter). This relationship reveals that for large errors (when noise dominates the signal), the correlation tends to zero. We can correct for this bias by applying a correction factor which removes the error terms in the above equation,
\begin{equation}
    r_{x,y}' = \left[\left(\frac{\var{x}-\sigma_x^2}{\var{x}}\right)\left(\frac{\var{y}-\sigma_y^2}{\var{y}}\right)\right]^{-1/2} r_{x,y} \,. \label{eq:correlation-corrected}
\end{equation}

\subsection{HR and Age Correlation}
\label{sec:correlation/local}

For the case of HR and Age, only for HR can we separate its uncertainties. This changes \autoref{eq:correlation-biased} and \autoref{eq:correlation-corrected} to become

\begin{align}
    r_{\text{HR},A}  & = \frac{\text{Cov}[\text{HR}^*, A^*] + \sigma^2_\text{HR, A}}{\sqrt{\left(\var{\text{HR}^*} + \sigma_\text{HR}^2 \right) \var{A}}} \label{eq:correlation-biased-2}\,, \\
    r'_{\text{HR},A} & = \left(\frac{\var{\text{HR}}-\sigma_\text{HR}^2}{\var{\text{HR}}}\right)^{-1/2} r_{\text{HR},A} \,, \label{eq:correlation-corrected-2}
\end{align}
respectively, where $A$ denotes Age.

Since we only have a sample of all these variables ($\text{HR}$ is the observed sample and $A$ is the posterior sample), we can only estimate the \emph{sample} correlation (the general calculation of which we defer to Appendix \ref{appendix:1}). We determine the biased sample correlation to be \num{-0.33 \pm 0.043} with significance $3.9\sigma$ and the corrected sample correlation to be \num{-0.37 \pm 0.047} with significance $4.0\sigma$, where the uncertainty in the estimate is derived from the variance of bootstrap samples. The bootstrap samples are generated by randomly sampling 102 rows (i.e., SNe) with replacement, estimating the correlation using the same technique in each case, and repeating this to get 100 correlation estimates.
The same procedure is applied with the global Age in place of the local Age. We determine the biased sample correlation to be \num{-0.32 \pm 0.070} with significance $3.3\sigma$ and the corrected sample correlation to be \num{-0.36 \pm 0.078} with significance $3.4\sigma$.

\RoseMC calculates the Spearman correlation coefficient instead of the Pearson coefficient we have presented herein. Unfortunately, we cannot use the same correction factor to estimate the Spearman correlation coefficient as it requires a nontrivial estimation of the variance of the rank statistics for a sample of independent, nonidentical Gaussian-distributed random variables. However, we do attempt to make a better estimate than \RoseMC using MC simulations on the HR and Age posterior samples\footnote{\RoseMC do not provide, in detail, how they estimated the Spearman correlation coefficient. We are able to reproduce their Spearman values with less than 5\% error by using the HR values without errors for every SN to every value in the Age posterior samples. We found that the ``dense'' rank statistics --- repeated values in the sample are assigned the same rank --- better match the \RoseMC results.}. Unlike \RoseMC, we account for the variability in HR by sampling under their (assumed) Gaussian distribution. Our simulation results in a Spearman correlation coefficient of \num{-0.255 \pm 0.091} with significance $2.5\sigma$ and \num{-0.245 \pm 0.084} with significance $2.5\sigma$ using local and global Age, respectively. These values are greater (in an absolute sense) and more significant than the \RoseMC estimates, albeit still insignificant at a $3\sigma$ threshold. However, we caution that we cannot confirm if our MC simulation is unbiased as we did for the Pearson coefficient.


\section{Slope}
\label{sec:slope}

As previously stated, the estimation of the parameters of a linear relationship between HR and Age (e.g., slope) is independent from the estimation of correlation owing to the necessary assumption that there indeed exists a linear relationship (forcing the correlation coefficient to be 1). Here we make that assumption and examine the resulting slopes.

\subsection{Models}
\label{sec:slope/models}

We compare the slope estimates from five models: ordinary least squares (OLS), orthogonal distance regression (ODR), LINMIX (model from \citealt{Kelly2007}; results taken from \Lee), direct estimation using posterior samples (results taken from \RoseMC), and our proposed model.
All models share the Bayesian linear regression form,
\begin{equation}
    y_i = \slope x_i^* + \intercept + \epsilon_\text{scatter} \,, \label{eq:linear-model}
\end{equation}

\noindent with $\slope$ and $\intercept$ being the slope and intercept (respectively), the asterisks denoting true variables, and the $\epsilon_\text{scatter}$ being the intrinsic scatter term.

Sharing only \autoref{eq:linear-model}, the models differ in their assumption about the errors in the observed values of HR $y_i$ (no asterisk) and true Age value $x^*_i$ for each SN: (1) OLS is a naive model that ignores all errors in both variables for each observation, (2) ODR assumes classical Gaussian errors in both variables, (3) direct estimation ignores errors in HR and associates every value of HR with every value in the Age posterior sample, (4) LINMIX assumes Gaussian errors in each observation of HR $y_i$ but assumes the \emph{entire population} of SNe Ages $x^*$ has the GMM distribution, and (5) our proposed model assumes Gaussian errors in HR and that each SN's Age $x^*_i$ (notice the subscript $i$) has the GMM distribution (see Sec.~\ref{sec:slope/proposed-model} for more details).

\subsection{Our Proposed Model}
\label{sec:slope/proposed-model}

We propose our own Bayesian model to apply a principled approach to make statistical inference using a Bayesian posterior as described in Section~\ref{sec:age_posterior_inference}. This approach rectifies two major issues: (1) the underestimation of the slope due to uncertainties (regression dilution), and (2) incorrect probability density distribution assumed for the Age posterior samples as apparent in \Lee. Our proposed model is composed of the following linear model, likelihood, and prior components:

\begin{equation*}
    \begin{aligned}
    \text{HR}_i             & = \text{HR}_i^* + \epsilon_\text{HR,i}                              \\
    \text{HR}_i^*           & = \slope A_i^* + \intercept + \epsilon_\text{scatter}               \\
    \epsilon_\text{HR,i}    & \sim \text{Normal}\left(0,\, \sigma_{\text{HR},i}\right)            \\
    \epsilon_\text{scatter} & \sim \text{Normal}(0, \sigma_\text{scatter})\,                      
    \end{aligned}
    \quad
    \begin{aligned}
    A_i^*                   & \sim \text{GMM}(\vec{w}_i,\, \vec{\upsilon}_i,\, \vec{\tau}_i;\, k) \\
    \intercept              & \sim \text{Uniform}(-1,\, 1)                                        \\
    \slope                  & \sim \text{Uniform}(-1,\, 0)                                        \\
    \sigma_\text{scatter}   & \sim \text{HalfNormal}(2)\,,
    \end{aligned}
\end{equation*}

\noindent where the tilde symbol ``$\sim$'' denotes that the left-hand side is distributed as the right-hand side, $A^*_i$ has an informative prior modeled after its posterior GMM fitted distribution with $\vec{w}_i$ being the GMM weight vector, $\vec{\upsilon}_i$ being the GMM mean vector, and $\vec{\tau}_i$ being the GMM standard deviation vector for the given $i$-th observation of vector size $k=3$ (the number of components in our employed GMM). Parameters $\intercept$ and $\slope$ are assumed with top-hat priors while $\sigma_\text{scatter}$ is a latent variable with uninformative half-normal prior since the parameter is nonnegative. We have set the scale parameter in the half-normal prior to be 2 such that it is sufficiently large for about 95\% of its values to be $< 4$.

\begin{figure}
    \centering
    \begin{subfigure}{0.5\textwidth}
        \centering
        \includegraphics[width=\linewidth]{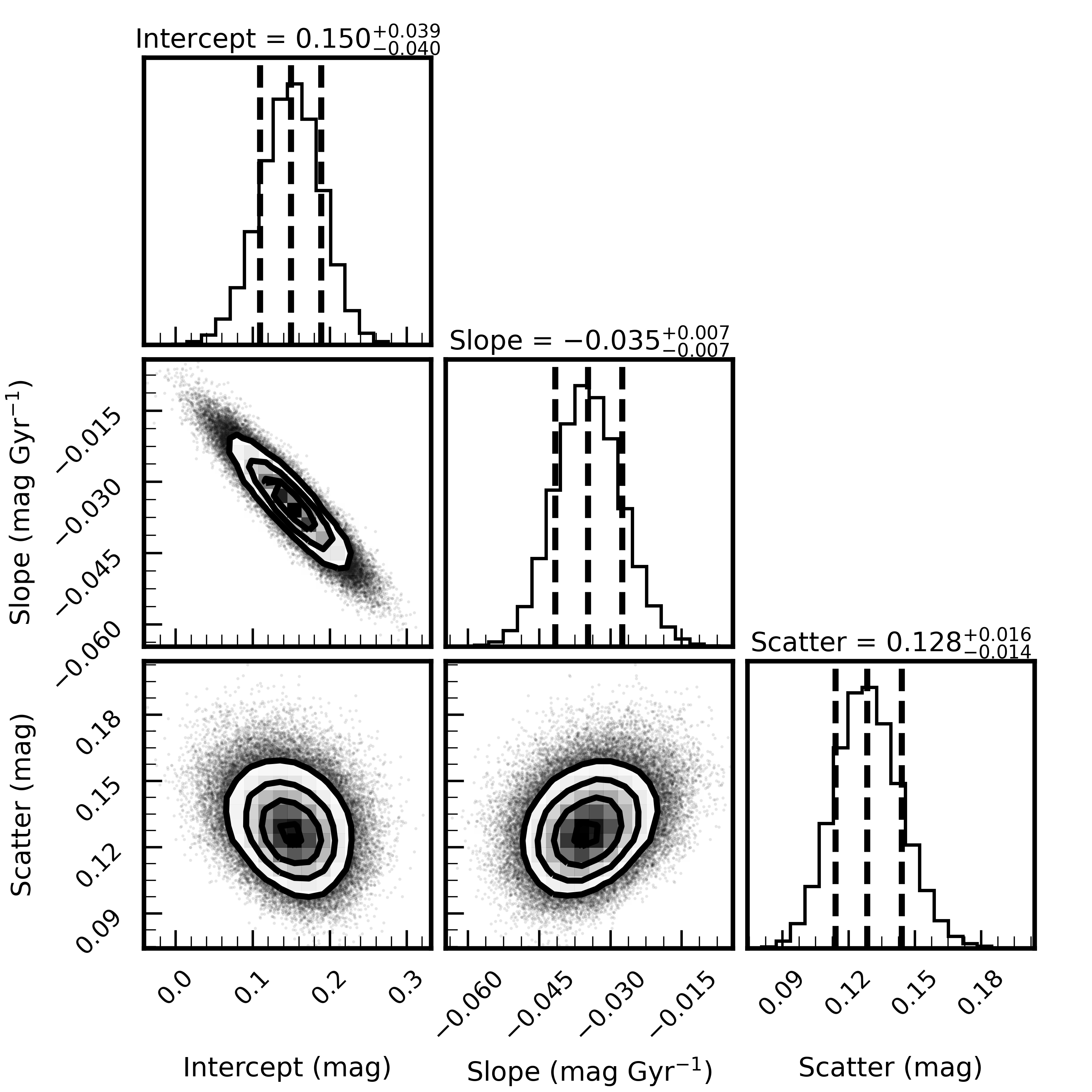}
    \end{subfigure}\hfill%
    \begin{subfigure}{0.5\textwidth}
        \centering
        \includegraphics[width=\linewidth]{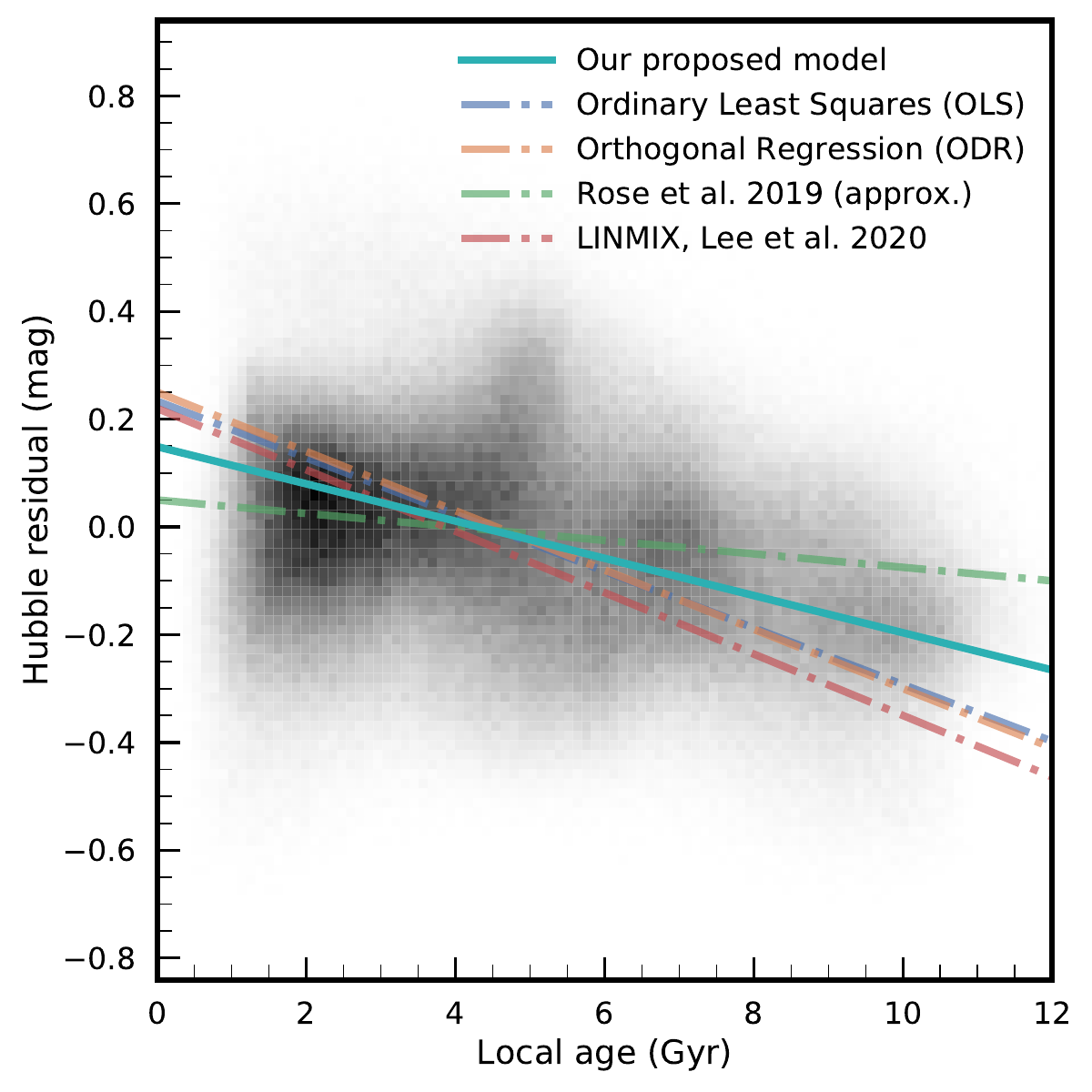}
    \end{subfigure}
    \caption{{\it (Top)} MCMC corner plot for the linear-regression parameters between HR and local Age for the proposed model described in Section~\ref{sec:slope/proposed-model}. {\it (Bottom)} Line fit using the models described in Section~\ref{sec:slope/models} and its parameters recorded in Table~\ref{tab:summary}. The background two-dimensional histogram shows the density of points in the HR with Gaussian noise from its uncertainty and Age with random values taken from its posterior samples.}
    \label{fig:corner-plot}
\end{figure}

The proposed model estimates the slope using MCMC. We use the pyMC3 NUTS \citep{Salvatier2016} implementation resulting in a slope and intercept posterior sample of size \num{100000}, then \num{76000} after burn-in as shown in Figure~\ref{fig:corner-plot}. The posterior samples for both parameters are nearly Gaussian with slight differences in the upper and lower uncertainties. We report the point estimate of the intercept is $0.080 \pm 0.035$\,mag and the slope is $-0.030 \pm 0.010$\,mag\,Gyr$^{-1}$.

\subsection{HR and Age Slope Estimations}
\label{sec:slope/local}
Under the assumption that there exists a linear relation between HR and local Age, our proposed model yields a slope of $-0.035 \pm 0.007$\,mag\,Gyr$^{-1}$ and an intercept of $0.151 \pm 0.04$\,mag.
With the same procedure applied for global Age, our proposed model yields a fitted slope of $-0.036 \pm 0.007$\,mag\,Gyr$^{-1}$ and an intercept of $0.16 \pm 0.04$\,mag.

\begin{table*}
    \caption{Linear regression parameter estimates for HR vs. local and global Age}
    \label{tab:summary}
    \begin{tabular}{lrrrr|rrrr}
        \hline
        {}                                    & \multicolumn{4}{c|}{Local} & \multicolumn{4}{c}{Global}                                                                                                                                                 \\
        {}                                    & slope                      & $\sigma_\mathrm{slope}$    & intercept   & $\sigma_\mathrm{intercept}$ & slope                     & $\sigma_\mathrm{slope}$   & intercept   & $\sigma_\mathrm{intercept}$ \\
        {}                                    & (\si{\mag\per\giga\year})  & (\si{\mag\per\giga\year})  & (\si{\mag}) & (\si{\mag})                 & (\si{\mag\per\giga\year}) & (\si{\mag\per\giga\year}) & (\si{\mag}) & (\si{\mag})                 \\
        \hline
        OLS                                   & $-0.053$                   &                            & 0.233       &                             & $-0.040$                  &                           & 0.178       &                             \\
        ODR                                   & $-0.055$                   & 0.014                      & 0.250       & 0.080                       & $-0.051$                  & $-0.012$                  & 0.260       & 0.070                       \\
        Rose et al. (2019)\textsuperscript{a} & $-0.012$                   &                            & 0.050       &                             & $-0.011$                  &                           & 0.050       &                             \\
        Lee et al. (2020)\textsuperscript{b}  & $-0.057$                   & 0.016                      & 0.220       &                             & $-0.047$                  & 0.011                     & 0.200       &                             \\
        Proposed Model                        & $-0.035$                   & 0.007                      & 0.151       & 0.040                       & $-0.036$                  & 0.007                     & 0.162       & 0.039                       \\
        \hline
    \end{tabular}\\
    \textsuperscript{a} Estimated from \citet[Fig. 5 and Fig. 6]{Rose2019};
    \textsuperscript{b} Slope taken and intercept estimated from \citet[Fig. 1]{Lee2020};\\
    Not all models give a standard deviation ($\sigma$) of the estimate.
\end{table*}


\section{Discussion and Conclusion}
\label{sec:conclusion}
We have reanalysed the results of \RoseMC and \Lee for the correlation and linear-model parameter estimates (respectively) that describe the relationship between HR and SN~Ia local and global environment ages. Our estimates properly account for uncertainties in the HR and the Age dataset that \emph{are} posterior samples produced by a Bayesian model in \RoseMC and \citet{Campbell2013}. In stark contrast to \Lee, we do not assume the uncertainties in the Age dataset to be Gaussian-distributed (given the multimodality observed in the posterior samples for some SNe), and unlike \RoseMC, we do not directly use the Age posterior samples as if we observed all values in the Age posterior samples.

We compare our correlation coefficient estimates with those of \RoseMC. First, our Spearman correlation coefficient calculated with an MC simulation results in greater (in an absolute sense) and more significant Spearman values than \RoseMC. Although we cannot test whether our method fully accounts for biases due to uncertainties in the HR and Age dataset, we agree with \RoseMC that there exists no significant Spearman correlation relation between HR and Age (local and global) above $3\sigma$. The Spearman correlation does not paint the full picture of possible linear relation between HR and Age. As \RoseMC mentioned, the Spearman correlation gauges the monotonic relation while the Pearson correlation gauges the linear relation. However, it is possible that the Pearson correlation is stronger than the Spearman correlation and vice versa. Fortunately, for estimating the Pearson correlation coefficient, we can confirm Equation~\ref{eq:correlation-biased} and correct Equation~\ref{eq:correlation-corrected} for the bias due to uncertainties in the datasets. After applying this bias correction, we find significant Pearson correlations of $4.0\sigma$ and $3.4\sigma$ for HR with the respective local and global Age sample.

While the monotonic relation may be insignificant, the linear relation is indeed significant, motivating a linear model. We estimate linear model parameters (slope and intercept) assuming that there exists a linear relationship between Hubble residuals and Age. Classical methods of regression (e.g., ordinary least squares and orthogonal distance regression) do not fully account for the nature of the Bayesian model posteriors, especially non-Gaussian-distributed posteriors that \emph{are} the Age posterior samples. We feed our estimator (MCMC) not the posterior samples but instead a posterior distribution from fitting a three-component Gaussian mixture model on the Age posterior samples. This posterior distribution in our MCMC is treated as an informative prior --- a stepwise approach motivated from hierarchical Bayesian models. Our proposed model's estimated slopes are in strong disagreement with --- and in between --- those reported by both \RoseMC (our values are higher) and \Lee (our values are lower), as summarised in Table~\ref{tab:summary}.

The correlation between HR and Age is significant within our dataset, suggesting that current standardisation procedures do not fully account for the effects of host environment age. In this \textit{Letter}, we do not attempt to correct standardisation, and we discourage direct usage of our linear parameters (e.g., Age step) without accounting for correlations between age and the light-curve-shape parameters. Other models can be inferred from Figure~\ref{fig:corner-plot}; for example, instead of fitting a single slope to all of the data, the bulk of SNe~Ia at relatively young ages ($\lesssim\SI{5}{\giga\year}$) may suggest a constant HR while the relation for older SNe~Ia still has an appreciable slope. Alternatively, neither mass nor age can fully describe the relationship themselves and other host properties could be at play. In our separate work, \cite{Murakami2021L} found the separations in HR values, between host-galaxy morphology separated into early- and late- types, are also consistent with both the HR--age slope (from \RoseMC and this work) as well as the HR--mass slope (from in \cite{Uddin2020}). We encourage further experimentation with standardisation to correct for luminosity evolution.

\section*{Acknowledgements}
A.V.F.'s group at U.C. Berkeley acknowledges generous support from Marc J. Staley, the Christopher R. Redlich Fund, Sunil Nagaraj, Landon Noll, the TABASGO Foundation, and the Miller Institute for Basic Research in Science (U.C. Berkeley). We thank Saurabh Jha for noticing an error in our original version of Figure 2.


\section*{Data Availability}
\label{sec:data-availability}
The raw data used in our analysis will be shared upon request to an author of this paper.



\bibliographystyle{mnras}
\bibliography{main.bib}

\begin{thebibliography}{}
\makeatletter
\relax
\def\mn@urlcharsother{\let\do\@makeother \do\$\do\&\do\#\do\^\do\_\do\%\do\~}
\def\mn@doi{\begingroup\mn@urlcharsother \@ifnextchar [ {\mn@doi@}
  {\mn@doi@[]}}
\def\mn@doi@[#1]#2{\def\@tempa{#1}\ifx\@tempa\@empty \href
  {http://dx.doi.org/#2} {doi:#2}\else \href {http://dx.doi.org/#2} {#1}\fi
  \endgroup}
\def\mn@eprint#1#2{\mn@eprint@#1:#2::\@nil}
\def\mn@eprint@arXiv#1{\href {http://arxiv.org/abs/#1} {{\tt arXiv:#1}}}
\def\mn@eprint@dblp#1{\href {http://dblp.uni-trier.de/rec/bibtex/#1.xml}
  {dblp:#1}}
\def\mn@eprint@#1:#2:#3:#4\@nil{\def\@tempa {#1}\def\@tempb {#2}\def\@tempc
  {#3}\ifx \@tempc \@empty \let \@tempc \@tempb \let \@tempb \@tempa \fi \ifx
  \@tempb \@empty \def\@tempb {arXiv}\fi \@ifundefined
  {mn@eprint@\@tempb}{\@tempb:\@tempc}{\expandafter \expandafter \csname
  mn@eprint@\@tempb\endcsname \expandafter{\@tempc}}}

\bibitem[\protect\citeauthoryear{{Betoule} et~al.,}{{Betoule}
  et~al.}{2014}]{Betoule2014}
{Betoule} M.,  et~al., 2014, \mn@doi [\aap] {10.1051/0004-6361/201423413},
  \href {https://ui.adsabs.harvard.edu/abs/2014A&A...568A..22B} {568, A22}

\bibitem[\protect\citeauthoryear{{Campbell} et~al.,}{{Campbell}
  et~al.}{2013}]{Campbell2013}
{Campbell} H.,  et~al., 2013, \mn@doi [\apj] {10.1088/0004-637X/763/2/88},
  \href {https://ui.adsabs.harvard.edu/abs/2013ApJ...763...88C} {763, 88}

\bibitem[\protect\citeauthoryear{{Childress} et~al.,}{{Childress}
  et~al.}{2013}]{Childress2013}
{Childress} M.,  et~al., 2013, \mn@doi [\apj] {10.1088/0004-637X/770/2/108},
  \href {https://ui.adsabs.harvard.edu/abs/2013ApJ...770..108C} {770, 108}

\bibitem[\protect\citeauthoryear{{Childress}, {Wolf}  \& {Zahid}}{{Childress}
  et~al.}{2014}]{Childress2014}
{Childress} M.~J.,  {Wolf} C.,   {Zahid} H.~J.,  2014, \mn@doi [\mnras]
  {10.1093/mnras/stu1892}, \href
  {https://ui.adsabs.harvard.edu/abs/2014MNRAS.445.1898C} {445, 1898}

\bibitem[\protect\citeauthoryear{{Kang}, {Lee}, {Kim}, {Chung}  \&
  {Ree}}{{Kang} et~al.}{2020}]{Kang2020}
{Kang} Y.,  {Lee} Y.-W.,  {Kim} Y.-L.,  {Chung} C.,   {Ree} C.~H.,  2020,
  \mn@doi [\apj] {10.3847/1538-4357/ab5afc}, \href
  {https://ui.adsabs.harvard.edu/abs/2020ApJ...889....8K} {889, 8}

\bibitem[\protect\citeauthoryear{{Kelly}}{{Kelly}}{2007}]{Kelly2007}
{Kelly} B.~C.,  2007, \mn@doi [\apj] {10.1086/519947}, \href
  {https://ui.adsabs.harvard.edu/abs/2007ApJ...665.1489K} {665, 1489}

\bibitem[\protect\citeauthoryear{{Lampeitl} et~al.,}{{Lampeitl}
  et~al.}{2010}]{Lampeitl2010}
{Lampeitl} H.,  et~al., 2010, \mn@doi [\apj] {10.1088/0004-637X/722/1/566},
  \href {https://ui.adsabs.harvard.edu/abs/2010ApJ...722..566L} {722, 566}

\bibitem[\protect\citeauthoryear{{Lee}, {Chung}, {Kang}  \& {Jee}}{{Lee}
  et~al.}{2020}]{Lee2020}
{Lee} Y.-W.,  {Chung} C.,  {Kang} Y.,   {Jee} M.~J.,  2020, \mn@doi [\apj]
  {10.3847/1538-4357/abb3c6}, \href
  {https://ui.adsabs.harvard.edu/abs/2020ApJ...903...22L} {903, 22}

\bibitem[\protect\citeauthoryear{{Murakami}, {Stahl}, {Zhang}, {Chu},
  {McGinness}, {Patra}  \& {Filippenko}}{{Murakami}
  et~al.}{2021}]{Murakami2021L}
{Murakami} Y.~S.,  {Stahl} B.~E.,  {Zhang} K.~D.,  {Chu} M.~R.,  {McGinness}
  E.~C.,  {Patra} K.~C.,   {Filippenko} A.~V.,  2021, \mn@doi [\mnras]
  {10.1093/mnrasl/slab034}, \href
  {https://ui.adsabs.harvard.edu/abs/2021MNRAS.504L..34M} {504, L34}

\bibitem[\protect\citeauthoryear{{Perlmutter}, {Aldering}, {Goldhaber}
  et~al.}{{Perlmutter} et~al.}{1999}]{Perlmutter1999}
{Perlmutter} S.,  {Aldering} G.,  {Goldhaber} G.,   et~al., 1999, \mn@doi
  [\apj] {10.1086/307221}, \href
  {http://adsabs.harvard.edu/abs/1999ApJ...517..565P} {517, 565}

\bibitem[\protect\citeauthoryear{{Phillips}}{{Phillips}}{1993}]{Phillips1993}
{Phillips} M.~M.,  1993, \mn@doi [\apjl] {10.1086/186970}, \href
  {http://adsabs.harvard.edu/abs/1993ApJ...413L.105P} {413, L105}

\bibitem[\protect\citeauthoryear{{Riess}, {Filippenko}, {Challis}
  et~al.}{{Riess} et~al.}{1998}]{Riess1998}
{Riess} A.~G.,  {Filippenko} A.~V.,  {Challis} P.,   et~al., 1998, \mn@doi
  [\aj] {10.1086/300499}, \href
  {http://adsabs.harvard.edu/abs/1998AJ....116.1009R} {116, 1009}

\bibitem[\protect\citeauthoryear{{Rigault} et~al.,}{{Rigault}
  et~al.}{2020}]{Rigault2020}
{Rigault} M.,  et~al., 2020, \aap, \href
  {https://ui.adsabs.harvard.edu/abs/2018arXiv180603849R} {}

\bibitem[\protect\citeauthoryear{{Rose}, {Garnavich}  \& {Berg}}{{Rose}
  et~al.}{2019}]{Rose2019}
{Rose} B.~M.,  {Garnavich} P.~M.,   {Berg} M.~A.,  2019, \mn@doi [\apj]
  {10.3847/1538-4357/ab0704}, \href
  {https://ui.adsabs.harvard.edu/abs/2019ApJ...874...32R} {874, 32}

\bibitem[\protect\citeauthoryear{{Rose} et~al.,}{{Rose}
  et~al.}{2020}]{Rose2020}
{Rose} B.~M.,  et~al., 2020, \mn@doi [\apjl] {10.3847/2041-8213/ab94ad}, \href
  {https://ui.adsabs.harvard.edu/abs/2020ApJ...896L...4R} {896, L4}

\bibitem[\protect\citeauthoryear{{Sako} et~al.,}{{Sako}
  et~al.}{2008}]{Sako2008}
{Sako} M.,  et~al., 2008, \mn@doi [\aj] {10.1088/0004-6256/135/1/348}, \href
  {https://ui.adsabs.harvard.edu/abs/2008AJ....135..348S} {135, 348}

\bibitem[\protect\citeauthoryear{Salvatier, Wiecki  \& Fonnesbeck}{Salvatier
  et~al.}{2016}]{Salvatier2016}
Salvatier J.,  Wiecki T.~V.,   Fonnesbeck C.,  2016, \mn@doi [{PeerJ} Computer
  Science] {10.7717/peerj-cs.55}, 2, e55

\bibitem[\protect\citeauthoryear{{Scolnic} et~al.,}{{Scolnic}
  et~al.}{2018}]{Scolnic2018}
{Scolnic} D.~M.,  et~al., 2018, \mn@doi [\apj] {10.3847/1538-4357/aab9bb},
  \href {https://ui.adsabs.harvard.edu/abs/2018ApJ...859..101S} {859, 101}

\bibitem[\protect\citeauthoryear{{Sullivan} et~al.,}{{Sullivan}
  et~al.}{2010}]{Sullivan2010}
{Sullivan} M.,  et~al., 2010, \mn@doi [\mnras]
  {10.1111/j.1365-2966.2010.16731.x}, \href
  {https://ui.adsabs.harvard.edu/abs/2010MNRAS.406..782S} {406, 782}

\bibitem[\protect\citeauthoryear{{Uddin} et~al.,}{{Uddin}
  et~al.}{2020}]{Uddin2020}
{Uddin} S.~A.,  et~al., 2020, \mn@doi [\apj] {10.3847/1538-4357/abafb7}, \href
  {https://ui.adsabs.harvard.edu/abs/2020ApJ...901..143U} {901, 143}

\makeatother
\end{thebibliography}
\vfill


\appendix
\section{Components for Pearson Correlation Coefficient}
\label{appendix:1}

Here we define all three terms that were undefined in Equations~\ref{eq:correlation-biased-2} and \ref{eq:correlation-corrected-2}. All expected values ($\E{\cdot}$) mentioned below are estimated with the sample mean. First, the sample variance for HR is

\begin{equation}
    \var{\text{HR}} = \frac{1}{N-1}\sum_{i=1}^{N} \left(\E{\text{HR}} - \text{HR}_i \right)^2 \,,\quad \text{as } N \to \infty \,.
\end{equation}

The two remaining terms (the sample variance of $A$ and the sample covariance of $\text{HR}$ and $A$) are nontrivial. The sample variance of $A$ is calculated by relying on the law of total variance,

\begin{gather}
    \var{A} = \E{\var{A \mid \text{HR}}} + \var{\E{A \mid \text{HR}}}, \stepcounter{equation}\\
    \var{A} = \sum_{i=1}^{N} \left( \frac{\var{A_i}}{N} + \frac{\left[\E{A} - \E{A_i}\right]^2}{N-1} \right) \,,\quad \text{as } N \to \infty \,. \label{eq:age-sample-variance}
\end{gather}

\noindent Finally, the sample covariance of $\text{HR}$ and $A$,

\begin{gather}
    \text{Cov}\left[\text{HR}, A\right] = \E{\text{HR} \cdot A} + \E{\text{HR}} \cdot \E{A}, \\
    \text{Cov}\left[\text{HR}, A\right] = \left(\frac{1}{N}\sum_{i=1}^{N} \E{A_i} \cdot \text{HR}_i \right) + \E{\text{HR}} \cdot \E{A} \,,\quad \text{as } N \to \infty \,. \label{eq:age-sample-covariance}
\end{gather}




\bsp	
\label{lastpage}
\end{document}